\documentclass{emulateapj}
\usepackage{graphicx}
\usepackage{epsfig}

\newcommand{\be}{\begin{equation}}
\newcommand{\ee}{\end{equation}}
\newcommand{\ba}{\begin{eqnarray}}
\newcommand{\ea}{\end{eqnarray}}

\shorttitle{Bayesian Evidence from Nested Sampling}
\shortauthors{P. Mukherjee, D. Parkinson and A. R. Liddle}

\begin{document}

\title{A Nested Sampling Algorithm for Cosmological Model Selection}
\author{Pia Mukherjee, David Parkinson and Andrew R. Liddle}
\affil{Astronomy Centre, University of Sussex, Brighton, BN1 9QH, United 
Kingdom}

\begin{abstract}
The abundance of new cosmological data becoming available means that a
wider range of cosmological models are testable than ever before.
However, an important distinction must be made between {\it parameter
fitting} and {\it model selection}.  While parameter fitting simply
determines how well a model fits the data, model selection statistics,
such as the Bayesian Evidence, are now necessary to choose between
these different models, and in particular to assess the need for new
parameters.  We implement a new evidence algorithm known as {\it
nested sampling}, which combines accuracy, generality of application
and computational feasibility, and apply it to some cosmological
datasets and models.
We find that a five-parameter model with 
Harrison--Zel'dovich initial spectrum is currently preferred.
\end{abstract}

\keywords{cosmology: theory}

\section{Introduction}

The higher-level inference problem of allowing the data to decide the
set of parameters to be used in fitting is known as {\it model
selection}.  In the Bayesian framework, the key model selection
statistic is the {\it Bayesian evidence} (Jeffreys 1961; MacKay 2003),
being the average likelihood of a model over its prior parameter
space. The evidence can be used to assign probabilities to models and
to robustly establish whether the data require additional parameters.

While the use of Bayesian methods is common practice in cosmological
parameter estimation, its natural extension to model selection has
lagged behind. Work in this area has been hampered by difficulties in
calculating the evidence to high enough accuracy to distinguish
between the models of interest.  The issue of model selection has been
raised in some recent papers (Marshall, Hobson \& Slosar 2003;
Niarchou, Jaffe \& Pogosian 2004; Saini, Weller \& Bridle 2004;
Bassett, Corasaniti \& Kunz 2004), and information criterion based
approximate methods introduced in Liddle (2004). Semi-analytic
approximations such as the Laplace approximation, which works well
only for gaussian likelihoods, and the Savage--Dickey density ratio
which works for more general likelihood functions but requires the
models being compared to be nested, are methods that were recently
exploited by Trotta (2005). A more general and accurate numerical
method is thermodynamic integration, but Beltr\'an et al.~(2005) found
that in realistic applications around $10^7$ likelihood evaluations
were needed per model to obtain good accuracy, making it a
supercomputer-class problem in cosmology where likelihood evaluations
are computationally costly.

In this paper we present a new algorithm which, for the first time,
combines accuracy, general applicability, and computational
feasibility. It is based on the method of {\em nested sampling},
proposed by Skilling (2004), in which the multi-dimensional integral
of the likelihood of the data over parameter space is performed using
Monte Carlo sampling, working through the prior volume to the regions
of high likelihood.

\vspace{3mm}

\section{Bayesian Inference}

Using Bayes' theorem, the probability that a model (hypothesis: $H$)
is true in light of observed data ($D$) is given by
\begin{equation}
P(H|D)=\frac{P(D|H)P(H)}{P(D)} \,.
\label{BayesTheorem}
\end{equation}
It shows how our prior knowledge $P(H)$ is modified in the presence of
data.

The posterior probability of the parameters ($\theta$) of a model in
light of data is given by
\begin{equation}
P(\theta|D,H)=\frac{P(D|\theta,H)P(\theta|H)}{P(D|H)}\,,
\end{equation}
where $P(D|\theta,H)$ is the likelihood of the data given the model
and its parameters, and $P(\theta|H)$ is the prior on the
parameters. This is the relevant quantity for parameter estimation
within a model, for which the denominator $P(D|H)$ is of no
consequence. $P(D|H)$ is however the {\em evidence} for the model $H$,
the key quantity of interest for the purpose of model selection
(Jeffreys 1961; MacKay 2003; Gregory 2005). Normalizing the posterior
$P(\theta|D,H)$ marginalized over $\theta$ to unity, it is given by
\begin{equation}
E=P(D|H)=\int d\theta \, P(D|\theta,H)P(\theta|H)\,,
\label{evdef}
\end{equation}
the prior also being normalized to unity. 

The evidence for a given model is thus the normalizing constant that
sets the area under the posterior $P(\theta|D,H)$ to unity. In the
now-standard Markov Chain Monte Carlo method for tracing the posterior
(Gilks et al.~1996; Lewis \& Bridle 2002), the posterior is reflected
in the binned number density of accumulated samples. The evidence
found in this way would not generally be accurate as the algorithm
would sample the peaks of the probability distribution well, but would
under-sample the tails which might occupy a large volume of the prior.

When comparing two different models using Bayes Theorem, the ratio of
posterior probabilities of the two models would be the ratio of their
evidences (called the Bayes Factor) multiplied by the ratio of any
prior probabilities that we may wish to assign to these models
(Eq.~1). It can be seen from Eq.~(\ref{evdef}) that while more complex
models will generally result in better fits to the data, the evidence,
being proportional to the volume occupied by the posterior relative to
that occupied by the prior, automatically implements Occam's razor. It
favours simpler models with greater predictive power provided they
give a good fit the data, quantifying the tension between model
simplicity and the ability to fit to the data in the Bayesian sense.
Jeffreys (1961) provides a useful guide to what constitutes a
significant difference between two models: $1<\Delta \ln E<2.5$ is
substantial, $2.5<\Delta \ln E<5$ is strong, and $\Delta \ln E>5$ is
decisive. For reference, a $\Delta \ln E$ of $2.5$ corresponds to odds
of about 1 in 13.

While for parameter fitting the priors become irrelevant once the data
are good enough, for model selection some dependence on the prior
ranges always remains however good the data.  The dependence on prior
parameter ranges is a part of Bayesian reasoning, and priors should be
chosen to reflect our state of knowledge about the parameters before
the data came along. The Bayesian evidence is unbiased, as opposed to
approximations such as the information criteria.

Perhaps the most important application of model selection is in
assessing the need for a new parameter, describing some new physical
effect proposed to influence the data. Frequentist-style approaches
are commonplace, where one accepts the parameter on the basis of a
better fit, corresponding to an improvement in $\Delta(\ln L)$ by some
chosen threshold (leading to phrases such as `two-sigma
detection'). Such approaches are non-Bayesian: the evidence shows that
the size of the threshold depends both on the properties of the
dataset and on the priors, and in fact the more powerful the dataset
the higher the threshold that must be set (Trotta 2005). Further, as
the Bayesian evidence provides a rank-ordered list of models, the need
to choose a threshold is avoided (though one must still decide how
large a difference in evidence is needed for a robust conclusion).

The main purpose of this paper is to present an algorithm for evidence
computation with widespread applications. However as a specific
application we examine the need for extra parameters against the
simplest viable cosmological model, a $\Lambda$CDM model with
Harrison--Zel'dovich initial spectrum, whose five parameters are the
baryon density $\Omega_{{\rm b}} h^2$, cold dark matter density
$\Omega_{{\rm cdm}}h^2$, the Hubble parameter $H_0=100h {\rm
kms}^{-1}{\rm Mpc}^{-1}$ (or the ratio $\Theta$ of the approximate
sound horizon at decoupling to its angular diameter distance), the
optical depth $\tau$, and the amplitude $A_{{\rm s}}$ of primordial
perturbations. We study the case for two additional parameters, the
scalar spectral index and the dark energy equation of state (assumed
constant).

\section{Nested Sampling}

\subsection{Basic Method}

\begin{figure}
\epsscale{1.}
\plotone{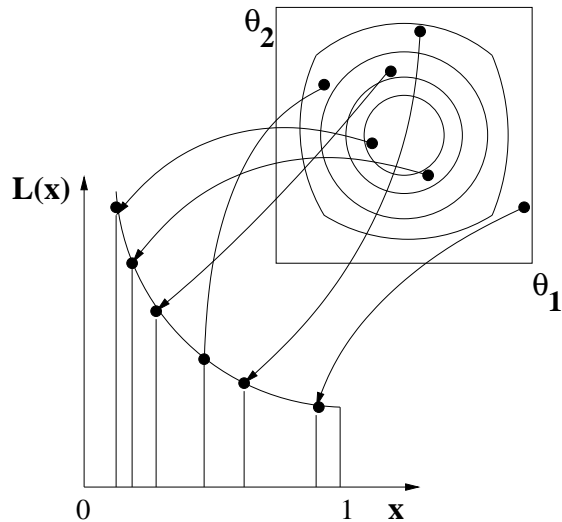}
\caption[nested]{\label{nested} The nested sampling algorithm
integrates the likelihood over the prior volume by peeling away thin
iso-surfaces of equal likelihood.
}
\end{figure}

Nested sampling (Skilling 2004) is a scheme to trace the variation of
the likelihood function with prior mass, with the effects of topology,
dimensionality and everything else implicitly built into it. It breaks
up the prior volume into a large number of `equal mass' points and
orders them by likelihood.  Rewriting Eq.~(\ref{evdef}) in the
notation of Skilling (2004), with $X$ as the fraction of total prior
mass such that $dX=P(\theta|H)d\theta$ and the likelihood
$P(D|\theta,H) = L(X)$, the equation for the evidence becomes
\begin{equation}
E = \int_0^1 L \, dX\,.  \label{Skillingdef} 
\end{equation} 
Thus the problem of calculating the evidence has become a
one-dimensional integral, in which the integrand is positive and
decreasing. Suppose we can evaluate the likelihood as $L_j = L(X_j)$,
where the $X_j$ are a sequence of decreasing values, such that
\begin{equation}
0 < X_m < ... < X_2 < X_1 < 1\,,
\end{equation}
as shown schematically in Figure~\ref{nested}.
Then the evidence can be estimated by any numerical method, for
example the trapezoid rule
\begin{equation}
E = \sum_{j=1}^m E_j\,,~~~~E_j=\frac{L_j}{2}(X_{j-1}-X_{j+1}) \,.
\end{equation}

The nested sampling algorithm achieves the above summation in the
following way:
\begin{enumerate}
\item Sample $N$ points randomly from within the prior, and evaluate their 
likelihoods. Initially we will have the full prior
range available, i.e.  $(0,X_0=1)$.

\item Select the point with the lowest likelihood $(L_j)$. The prior
volume corresponding to this point, $X_j$, can be estimated
probabilistically.  The average volume decrease is given as
$X_j/X_{j-1} = t$ where $t$ is the expectation value of the largest of
$N$ random numbers from uniform(0,1), which is $N/(N+1)$.

\item Increment the evidence by $E_j=L_j(X_{j-1}-X_{j+1})/2$.

\item Discard the lowest likelihood point and replace it with a new
point, which is uniformly distributed within the remaining prior
volume $(0,X_j)$. The new point must satisfy the hard constraint on
likelihood of $L>L_j$.

\item Repeat steps 2--4, until the evidence has been estimated to some
desired accuracy.
\end{enumerate}

\begin{figure}
\epsscale{1.15}
\plotone{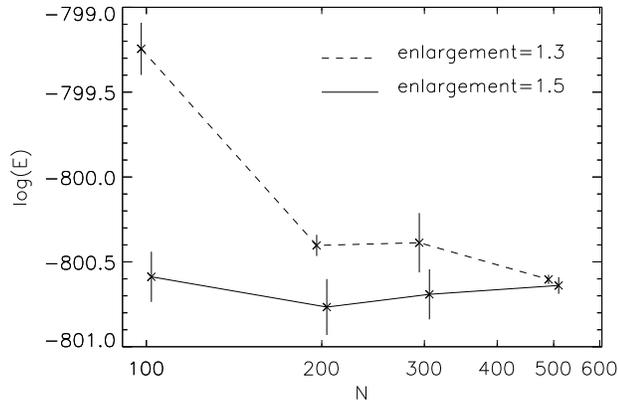}
\caption[nliveplot]{\label{nliveplot} Evidence against the number of
sample points $N$ for a 6D Gaussian likelihood
function as a test case. The horizontal lines show the actual
analytical value.Points are displaced slightly horizontally for visual clarity.
}
\end{figure}

The method thus works its way up the likelihood surface, through
nested surfaces of equal likelihood. After $j$ steps the prior mass
remaining shrinks to $X_j \sim (N /(N+1))^j$. The process is
terminated when some stopping criterion is satisfied, and a final
amount of $\langle L\rangle \times X_j$ due to the $N-1$ remaining
sample points is added to the thus far accumulated evidence.

Thus a multi-dimensional integral is performed using Monte Carlo
sampling, imposing a hard constraint in likelihood on samples that are
uniformly distributed in the prior, implying a regularized
probabilistic progression through the prior volume. Besides
implementing and testing this scheme in the cosmological context, our
main contribution lies in developing a general strategy to sample new
points efficiently.

\subsection{Details}

The prior space to sample from reduces by a constant factor of N/(N+1)
every time the lowest likelihood point is discarded. The most
challenging task in implementing the algorithm is to sample uniformly
from the remaining prior volume, without creating too large an
overhead in likelihood evaluations even when the remaining volume of
prior space may be very small.  The new point must be uncorrelated to
the existing points, but we can still use the set of existing points
as a guide. We find the covariance of the live points, rotate our
coordinates to the principal axes, and create an ellipsoid which just
touches the maximum coordinate values of the existing points. To allow
for the iso-likelihood contours not being exactly elliptical, these
limits are expanded by a constant enlargement factor, aiming to
include the full volume with likelihood exceeding the current limit
(if this is not done new points will be biased towards the centre,
thus overestimating the evidence). New points are then selected
uniformly within the expanded ellipse until one has a likelihood
exceeding the old minimum, which then replaces the discarded
lowest-likelihood point.

\begin{figure}
\epsscale{1.15}
\plotone{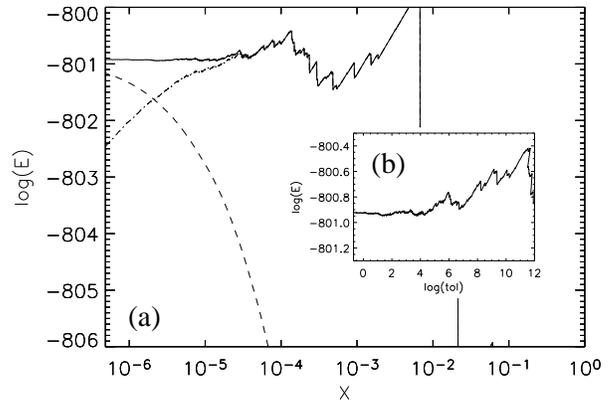}
\caption[stoppingcriterion]{\label{stoppingcriterion} (a) The
accumulated evidence (dashed curve), the evidence contributed by the
remaining points at each stage (dot-dashed curve), and their sum
(solid curve), shown against prior volume remaining for a 6D Gaussian
likelihood function. (b) A later part of the solid curve shown
against the log($tol$) (see text).}
\end{figure}

\begin{table*}
\centering
\caption{\label{table1} Parameter ranges and evidences for various
cosmological models. Other parameter ranges are given in text.}
\begin{tabular}{|c|c|c|c|c|c|}
\hline
Model & $\Lambda$CDM+HZ & $\Lambda$CDM+$n_{{\rm s}}$ &
$\Lambda$CDM+$n_{{\rm s}}$ & HZ+$w$ & $w+n_{{\rm s}}$\\   
& & & (wide prior) & & \\\hline
$n_{{\rm s}}$ & 1 & 0.8 -- 1.2 &  0.6 -- 1.4  & 1 & 0.8 -- 1.2 \\
\hline 
$w$ & -1 & -1 & -1 & -$\frac{1}{3}$ -- -1 & -$\frac{1}{3}$ -- -1\\ 
\hline \hline
e.f & 1.5 & 1.7 & 1.7 & 1.7 & 1.8 \\ \hline
$N_{\rm like} (\times 10^4)$ & 8.4 & 17.4 & 16.7 & 10.6 & 18.0 \\ \hline
$\log{E}$ & $0.00 \pm 0.08 $ & $-0.58 \pm 0.09$ & $-1.16 \pm 0.08$ & 
$-0.45 \pm 0.08$ & $-1.52 \pm 0.08$   \\ \hline
\end{tabular}
\end{table*}

The two algorithm parameters to be chosen are the number of points $N$
and the enlargement factor of the ellipsoid.  Figure 2 shows evidence
verses $N$, for a flat Harrison--Zel'dovich model with a cosmological
constant.  The mean evidence values and standard deviations obtained
from 4 repetitions are shown.  These are shown for two different
values of the enlargement factor.  When the enlargement factor is
large enough (here 1.5) the evidence obtained with 100 sample points
agrees with that obtained with 500, while when the enlargement factor
is not large enough, the evidences obtained with small N are
systematically biased high.  Similar tests done on multi-dimensional
Gaussian likelihood functions, for which the expected evidence can be
found analytically, indicated the same.  Based on such tests we choose
to work with $N$ of 300, and enlargement factors of 1.5 for the 5D
model (this corresponds to a 50\% increase in the range of each
parameter), 1.7 for 6D and 1.8 for 7D models.  These choices are
conservative (smaller values would reduce the computing time), and
were made in order to ensure evidences that are accurate and free of
systematics.  In our cosmological applications we have also computed
evidences with larger enlargement factors and found that they remained
unchanged.  The typical acceptance rate in finding a new point during
the course of the algorithm was found to be roughly constant at
$\sim20-25\%$ for an enlargement factor of 1.5, and lower for larger
enlargement factors, after an initial period of almost $100\%$
acceptance, as expected.

Figure 3a shows the accumulated evidence, the evidence contributed by
the remaining points at each stage, and their sum, against prior
volume remaining, again for a flat Harrison--Zel'dovich model with a
cosmological constant.  The $X$ at which the calculation can be
terminated will depend on the details of the problem
(e.g.~dimensionality, priors etc.).  We define a parameter $tol$ as
the maximum possible fractional amount that the remaining points could
increase the evidence by: $tol \equiv (L_{{\rm max}})_j X_j/E$ where
$L_{{\rm max}}$ is the maximum likelihood of the current set of sample
points.  Figure 3b zooms into a late part of the solid curve, now
plotting it against the value of the parameter $tol$.  The calculation
need only be carried out until the standard error on the mean
evidence, computed for a certain number of repetitions, drops below
the desired accuracy.  An uncertainty in $\ln E$ of 0.1 would be the
highest conceivable accuracy one might wish, and with 8 repetitions
this happens when $\ln(tol)\sim$ a few.  This takes us to quite small
$X$, of order $10^{-6}-10^{-7}$ in our actual cosmological
simulations.


\section{Results}

We have calculated the evidences of four different cosmological
models: 1) a flat, Harrison--Zel'dovich model with a cosmological
constant ($\Lambda$CDM+HZ), 2) the same as 1, except allowing the tilt
of the primordial perturbation spectrum $n_{{\rm s}}$ to vary
($\Lambda$CDM+$n_{{\rm s}}$), 3) the same as 1, except allowing the
equation of state of the dark energy to take alternative values to
$w=-1$ ($w$+HZ), and finally 4) allowing both $n_{{\rm s}}$ and $w$ to
vary ($w+n_{{\rm s}}$).  The prior ranges for the other parameters
were fixed at $0.018 \le \Omega_{{\rm b}} h^2 \le 0.032$, $0.04 \le
\Omega_{{\rm cdm}} h^2 \le 0.16$, $0.98 \le \Theta \le 1.1$, $0 \le
\tau \le 0.5$, and $2.6 \le \ln(A_{{\rm s}} \times 10^{10}) \le 4.2$.

The data sets we use are CMB TT and TE anisotropy power spectrum data
from the WMAP experiment (Bennett et al.~2003; Spergel et al.~2003;
Kogut et al.~2003), together with higher $l$ CMB temperature power
spectrum data from VSA (Dickinson et al.~2004), CBI (Pearson et
al.~2003) and ACBAR (Kuo et al.~2004), matter power spectrum data from
SDSS (Tegmark et al.~2004) and 2dFGRS (Percival et al.~2001), and
supernovae apparent magnitude--redshift data from Riess et al.~(2004).

Results are shown in Table 1. For the spectral tilt, evidences have
been found for two different prior ranges, as an additional test of
the method. For a prior range twice the size of the original in $n_s$
the evidence is expected to change by $-\ln 2$ at most and that
difference is recovered.  The first 2 rows show the priors on the
additional parameters of the model, or their constant values if they
were fixed.  The 3rd row shows the enlargement factor (e.f) we used
for the model. The 4th row shows the total number of likelihood
evaluations needed to compute the mean $\ln$ evidence to an accuracy $
\sim 0.1$, and the 5th row shows the mean $\ln E$ and the standard
error in that mean computed from 8 repetitions of the calculation,
normalized to the $\Lambda$CDM+HZ evidence.

The $\Lambda$CDM+HZ model has the highest evidence, and as such is the
preferred fit to the data. Hence we do not find any indication of a
need to introduce parameters beyond the base set of 5, in agreement
with the conclusions of Liddle (2004) and Trotta (2005).  However, the
difference between the $\ln E$ of the higher-dimensional models and
the base model is not large enough to significantly exclude any of those
 models at present.

\section{Conclusions}

We introduce the nested sampling algorithm for the computation of
Bayesian evidences for cosmological model selection. We find that this
new algorithm uniquely combines accuracy, general applicability and
computational feasibility. It is able to attain an accuracy (standard
error in the mean $\ln$ evidence) of 0.1 in O($10^5$) likelihood
evaluations. It is therefore much more efficient than thermodynamic
integration, which is the only other method that shares the general
applicability of nested sampling. Nested sampling also leads to a good
estimate of the posterior probability density of the parameters of the
model for free, which we will discuss in a forthcoming paper. We also
plan to make a public release of the code in the near future.

Using nested sampling we have computed the evidence for the simplest
cosmological model, with a base set of 5 parameters, which provides a
good fit to current cosmological data. We have computed the evidence
of models with additional parameters --- the scalar spectral tilt, 
a constant dark
energy equation of state parameter, and both of these together. We
find that current data offer no indication of a need to add extra
parameters to the base model, which has the highest evidence amongst
the models considered.

\acknowledgments
The authors were supported by PPARC. We thank Martin Kunz, Sam Leach, Peter 
Thomas and especially John Skilling for helpful advice and comments. 
The analysis was performed on the UK national cosmology supercomputer 
(COSMOS) in Cambridge.

\end{document}